\documentclass[aps,prd,twocolumn,showpacs,amsmath,amssymb]{revtex4}

\usepackage{amssymb}
\usepackage{mathrsfs}
\usepackage{txfonts}

\usepackage{graphicx}
\usepackage{dcolumn}% Align table columns on decimal point
\usepackage{bm}% bold math

\begin{document}

\title{Two models unifying warm inflation with dark matter and dark energy}

\author{Xiao-Min Zhang}
\email{zhangxm@mail.bnu.edu.cn}
\affiliation{School of Science, Qingdao University of Technology, Qingdao 266033, China}
\author{Kai Li}
\affiliation{School of Science, Qingdao University of Technology, Qingdao 266033, China}
\author{Yi-Fu Guo}
\affiliation{School of Science, Qingdao University of Technology, Qingdao 266033, China}
\author{Peng-Cheng Chu}
\email{kyois@126.com}
\affiliation{School of Science, Qingdao University of Technology, Qingdao 266033, China}
\author{He Liu}
\email{liuhe@qut.edu.cn}
\affiliation{School of Science, Qingdao University of Technology, Qingdao 266033, China}
\author{Jian-Yang Zhu}
\affiliation{Department of Physics, Beijing Normal University, Beijing 100875, China}

\date{\today}
%---------------------------------------------------------------------------------------------
\begin{abstract}
Two models that unify warm inflation with dark matter and dark energy are proposed. In the models, a single scalar field is responsible for the early expansion of the universe through the process of dissipative warm inflation and then acts as both dark matter and dark energy in subsequent stages. The first model is based on a noncanonical field with the Lagrangian density $\mathcal{L}=F(X)-V(\phi)$, where the potential is dominant at the slow-roll inflationary epoch and negligible in subsequent stages. The second model takes advantage of a $k$-essence Lagrangian density having the coupled form $\mathcal{L}=F(X)V(\phi)$. For both models, equations of the evolution for the fields and observational constraints are presented, and an evolution law describing how the energy density $\rho$ and state parameter $w$ scale with the scale factor $a$ is obtained.

\end{abstract}
\maketitle

%------------------------------------------------------------------------------------------------------------
\section{\label{sec:level1}Introduction}

The Universe underwent a quasi-exponential expansion, namely inflation \cite{Guth1981,Linde1982,Bassett2006}, at a very early age. Inflation is a necessary supplement to the standard model of cosmology, solving horizon, flatness and monopole problems and providing seeds for the cosmological perturbations that generate large-scale structure. There are two candidates for inflation, namely standard inflation and warm inflation. Warm inflation was first proposed by A. Berera in 1995 \cite{BereraFang,Berera1995}. Warm inflation not only inherits the advantages of standard inflation but also has its own advantages. In warm inflation, the slow-roll conditions are more easily satisfied because of the thermal damping effect \cite{Ian2008,Campo2010,Zhang2014,ZhangZhu}. The warm inflation picture can slove the ``$\eta$ problem'' \cite{etaproblem,etaproblem1} and the overlarge amplitude of inflaton \cite{Berera2006,BereraIanRamos} that occur in standard inflation. The origins of cosmological density fluctuations are different in the two inflationary scenarios. Cosmological fluctuations mainly originate from thermal fluctuations in warm inflation \cite{Berera2000,Lisa2004,Taylor2000,Chris2009,BereraIanRamos}, whereas they originate from vacuum quantum fluctuations in standard inflation \cite{LiddleLyth,Bassett2006}. During warm inflation, the inflaton field is not isolated but instead interacts with other subdominated boson or fermion fields, such that radiation is produced in appreciable amounts. Hence, the reheating phase is no longer needed and the Universe can proceed smoothly into the Big Bang radiation-dominated phase. One advantage of warm inflation is that the inflaton does not need to be dissipated completely to reheat the Universe, and a relic of the inflaton field can survive at a later time to act as dark matter and dark energy.

Theoretical research and observations have shown that the Universe appears to comprise approximately $32\%$ matter, where most are dark matter which clusters and drives the formation of the large-scale structure in the Universe, and $68\%$ dark energy, which drives the late-time acceleration of the Universe \cite{PLANCK1}. The $\Lambda$CDM model is one of our most successful phenomenological models and the predictions of it fit to the Planck data well \cite{PLANCK1}. The nature of neither component is known with certainty, and it is thus interesting and reasonable to consider that a simpler model might be possible, where a single component acts as inflaton, dark matter and dark energy. Unified models have been proposed to unify dark energy and dark matter using a pure kinetic field, tachyon field or complex axion field \cite{Scherrer2004,Bilic2009,Brandenberger2021}. Now that it is known that accelerated expansion is a common feature of the very early and the very late Universe, it is plausible that a common mechanism is responsible for the two stages of expansion. Several models, such as quintessential inflation \cite{quintessential} and interacting model \cite{Dimopoulos2019}, have been constructed to explain inflation and dark energy using a single scalar field. Besides the above category of models, schemes that attempt to unify dark matter and inflation have been reported \cite{Lidsey2002,Zsembinszki2007}. In addition, attempts have been made to unify all three of inflation, dark matter and dark energy in \cite{Bose2009,Nilok2009,Liddle2008,Sa2020,Arbey2021}. It is interesting and attractive idea to unify inflation, dark matter and dark energy in a single theoretical picture. All previous relevant works unified the two dark components with cold inflation. In this paper, we attempt to unify inflation, dark matter and dark energy in light of the warm inflationary mechanism, which produces appreciable radiation naturally accompanied by the inflation and provides a ``graceful exit'' to the radiation-dominated epoch. The noncanonical unified models under the cold inflationary framework successfully yield exponential inflation, but suffer from the rapid reheating and the graceful-exit problems. These problems can be neatly solved in warm inflation, and warm inflation has a broader choice of the potential. In this paper, we show that two certain scalar fields with nonstandard kinetic terms can serve as a unified model of warm inflation, dark matter and dark energy. The first type is a scalar field with a noncanonical kinetic term and a dominating potential term, with the two terms being separable in the Lagrangian $\mathcal{L}=F(X)-V(\phi)$. The second type is a $k$-essence, with a Lagrangian density having coupled kinetic and potential terms $\mathcal{L}=F(X)V(\phi)$. The two fields first generate enough $e$-folds in the very early inflationary regime, and later naturally produce an energy density that scales like the sum of a nonrelativistic dust component with the equation of state $w=0$ and a cosmological-constant-like component.

The remainder of the paper is organized as follows. Section \ref{sec:level2} introduces the framework and main equations of the original warm inflationary scenario. Section \ref{sec:level3} and Section \ref{sec:level4} analyze the two types of noncanonical scalar model with a Lagrangian of separable or coupling form, respectively, to give the unified picture of warm inflation, dark matter and dark energy. Finally, we summarize our results and present discussions in Section \ref{sec:level5}.

\section{\label{sec:level2}Framework of warm inflation}
We now briefly introduce the theory of warm inflation, using the canonical scalar field as the inflaton. The scalar inflaton field is not isolated but interacts with other sub-dominated fields in the warm inflationary picture. A significant amount of radiation was produced constantly during the inflationary epoch owing to the interaction. The Universe is therefore hot with a non-zero temperature $T$.
The total matter action of the multi-component Universe in warm inflation is
\begin{equation}\label{action}
  S=\int d^4x \sqrt{-g}  \left[ \mathcal{L}_{\phi}+\mathcal{L}_R+\mathcal{L}_{int}\right],
\end{equation}
where $\mathcal{L}_{\phi}$ is the Lagrangian density of the inflaton field, $\mathcal{L}_R$ is the Lagrangian density of radiation fields, and $\mathcal{L}_{int}$ denotes the interaction between the scalar fields. In the Friedmann-Robertson-Walker (FRW) Universe, the mean inflaton field is homogeneous; i.e., $\phi=\phi(t)$. By varying the action with respect to the inflaton field and adopting the assumptions in warm inflation \cite{BereraFang,Berera1999,Zhang2014}, we obtain the evolution equation for the inflaton field:
\begin{equation}
\ddot{\phi}+(3H+\Gamma )\dot{\phi}+V_{eff,\phi}=0  \label{EOMphi}.
\end{equation}
Here, the parameter $H$ is the Hubble parameter obeying the Friedmann equation:
\begin{equation}\label{Friedmann}
  3H^2=8\pi G\rho.
\end{equation}
In Eq. (\ref{EOMphi}), $\Gamma$ is the dissipation coefficient, $V_{eff}$ is the effective potential acquired thermal correction, and the subscript $\phi$ denotes a derivative. The effective potential $V_{eff}$ is different from the zero-temperature potential in cold inflation, but the thermal correction is small \cite{Ian2008,Campo2010,Zhang2014}. For simplicity, $V_{eff}$ is rewritten as $V$ in the following. The dissipation term $\Gamma \dot{\phi}$, describing the dissipation effect of $\phi $ toward radiation \cite{BereraFang,Berera2006,Berera2000,BereraIanRamos}, is a thermal
damping term. $\Gamma $ is often considered a constant in some papers for simplicity \cite{Herrera2006,Xiao2011,Taylor2000}. A different form of $\Gamma $ can be obtained using concrete models of the interaction between the inflaton field and other fields \cite{MossXiong2006,MarGil2013,BereraIanRamos}. Additionally, $\Gamma$ has been found to more likely be a function of the inflaton field and even the temperature.

An important parameter in warm inflationary theory is the dissipation strength $r$, which describes the effectiveness of warm inflation. This dimensionless parameter $r$ is conventionally defined as
\begin{equation}\label{r}
  r=\frac{\Gamma}{3H}.
\end{equation}
We have $r\gg 1$ for the strong dissipation regime of warm inflation and $r\ll 1$ for the weak dissipation
regime of warm inflation.

The thermal dissipative effect of warm inflation is accompanied by the production of entropy. The entropy density is expressed as $s=-\partial f/\partial T$ in the field of thermodynamics. The free energy density $f=\rho -Ts$ is dominated by the potential during inflation, and we thus have $s\simeq -V_T$.

The total energy density of the multi-component Universe is
\begin{equation}
\rho =\frac 12\dot{\phi}^2+V(\phi ,T)+Ts.  \label{rho}
\end{equation}
and the total pressure is
\begin{equation}
p=\frac 12\dot{\phi}^2-V(\phi ,T).  \label{p}
\end{equation}
The energy-momentum conservation equation $\dot{\rho}+3H(\rho +p)=0$ combined with Eq. (\ref{EOMphi}) yields the entropy production equation:
\begin{equation}
T\dot{s}+3HTs=\Gamma \dot{\phi}^2.  \label{entropy}
\end{equation}
The thermal correction to the effective potential is sufficiently small (as ensured by the slow-roll conditions), so the above equation is equivalent to the equation for the radiation energy density producing $\dot{\rho}_r+4H\rho _r=\Gamma\dot{\phi}^2$.

The exact inflationary equations are difficult to solve, and thus, we often apply slow-roll approximations during inflation to drop the highest derivative terms in the evolution equations and Friedmann equation:
\begin{equation}
3H(1+r)\dot{\phi}+V_{\phi}=0,  \label{SRdotphi}
\end{equation}
\begin{equation}
Ts=r\dot{\phi}^2,  \label{SRTs}
\end{equation}
\begin{equation}
H^2=\frac{8\pi G}3 V,  \label{SRH}
\end{equation}
\begin{equation}
4H\rho _r=\Gamma \dot{\phi}^2.  \label{SRrho}
\end{equation}
The validity of the slow-roll approximations depends on the slow-roll conditions given by stability analysis \cite{Ian2008,Campo2010,Lisa2004,Zhang2014}. The slow-roll conditions are associated with important slow-roll parameters defined as
\begin{equation}
\epsilon =\frac{M_p^2}{2}\left(\frac{V_{\phi}}{V}\right) ^2, \eta =M_p^2\frac {V_{\phi \phi}}{V}, \beta
=M_p^2\frac{V_{\phi}\Gamma_{\phi}}{V\Gamma},
\end{equation}
where $M_p^2=\frac 1{8\pi G}$.
The slow-roll approximations hold when $\epsilon\ll 1+r$, $\eta\ll1+r$ and $\beta\ll1+r$. Any quantity of order $\epsilon/(1+r)$ is described as being a first-order small quantity in the slow-roll approximations.
There are two additional slow-roll parameters when the dissipation coefficient depends on temperature: $b=\frac {TV_{\phi T}}{V_{\phi}}$ and $c=\frac{T\Gamma_T}{\Gamma}$. The slow-roll validity requires $b\ll1$, $|c|<4$ \cite{Ian2008,Campo2010}.

When the slow-roll parameter becomes $\epsilon \simeq 1+r$, we have $\ddot{a}=0$, implying the end of the inflationary
phase. The number of $e$-folds in warm inflation is then given by
\begin{equation}
N(\phi)=\int Hdt=-\frac 1{M_p^2}\int_\phi ^{\phi _e}\frac V{V_\phi} (1+r)d\phi,  \label{efold}
\end{equation}
where the subscript $e$ denotes the end of inflation.

\section{\label{sec:level3}Noncanonical unified model with a Lagrangian density having separable kinetic and potential terms}
We now consider a type of scalar field with a noncanonical Lagrangian having separable kinetic and potential terms:
\begin{equation}\label{L1}
  \mathcal{L}=F(X)-V(\phi),
\end{equation}
where $X=\frac12g^{\mu\nu}\partial_{\mu}\phi\partial_{\nu}\phi$. The variable $X$ reduces to $X=\frac12\dot\phi^2$ in the flat FRW Universe, because the field is homogeneous.
The kinetic term of the field is chosen ad hoc as
\begin{equation}\label{F1}
  F(X)=KX+LX^{\frac12},
\end{equation}
where parameters $K$ and $L$ are constants. The parameter $K$ is dimensionless, and $L$ should have the dimension $[M]^2$, where $[M]$ denotes the dimension of mass. A proper noncanonical Lagrangian density should satisfy the conditions $\mathcal{L}_X\geq0$ and $\mathcal{L}_{XX}\geq0$ (where a subscript $X$ here denotes a derivative) to obey the null energy condition and the physical propagation of perturbations \cite{Franche2010,Bean2008}. The parameters in our model should then be restricted to $L\leq0$ and $K>0$. The potential is chosen as $V(\phi)=\frac12m^2\phi^2$ conventionally.
The pressure of the inflaton field is
\begin{equation}\label{pressure}
p=\mathcal{L}=KX+LX^{\frac12}-\frac12m^2\phi^2,
\end{equation}
whereas the energy density is given by
\begin{equation}\label{rho1}
\rho=2X\mathcal{L}_X-\mathcal{L}=KX+\frac12m^2\phi^2.
\end{equation}
The state parameter is $w=\frac p{\rho}=\frac{KX+LX^{\frac12}-\frac12m^2\phi^2}{KX+\frac12m^2\phi^2}$, and we have $w\rightarrow-1$ in the potential-dominated slow-roll inflationary regime.

Adopting warm inflationary assumptions and varying the total action in warm inflation, the evolution equation for the scalar field acting as the inflaton are obtained as
\begin{equation}\label{EOM1}
  (F_X+2XF_{XX})\ddot\phi+3HF_X(1+r)\dot\phi+V_{\phi}=0,
\end{equation}
where $r=\frac{\Gamma}{3HF_X}$ is the refined dissipation strength parameter in noncanonical warm inflation, which remains dimensionless \cite{Zhang2021}. Using the variable $X$, the evolution equation is rewritten as
\begin{equation}\label{EOM2}
  (F_X+2XF_{XX})\dot X+6HF_X(1+r)X+\sqrt{2X}V_{\phi}=0.
\end{equation}
In the slow-roll regime, the evolution equation (\ref{EOM1}) reduces to $3HF_X(1+r)\dot\phi+m^2\phi=0$, and the energy density of radiation can be expressed as $\rho_r=\frac32rF_XX$ using Eq. (\ref{SRrho}). The validity of slow-roll approximations in noncanonical warm inflation can be easily guaranteed by the slow-roll conditions: $\epsilon\ll F_X(1+r),~\beta\ll F_X(1+r),~\eta\ll F_X r,~~b\ll\frac1{1+r}$ \cite{Zhang2021}. Adopting the slow-roll approximations, the number of $e$-folds can be obtained:
\begin{eqnarray}\label{efold1}
  N&=&\int Hdt=\int_{\phi_e}^{\phi_i}\frac{V}{V_{\phi}}(1+r)F_X d\phi \nonumber\\ &\simeq&2\pi GF_X(1+r)(\phi_i^2-\phi_e^2)
  \nonumber \\ &=&\frac{4\pi GF_X}{m^2}(1+r)(V_i-V_e),
\end{eqnarray}
where the subscript ``i'' refers to beginning of inflation and ``e'' refers to the end. Considering the condition that when $\epsilon\simeq F_X(1+r)$, inflation ends, we get $V_i=\frac{m^2(N+\frac12)}{4\pi GF_X(1+r)}$. As well as $N\simeq60$, our noncanonical warm inflation can sustain enough time to solve the horizon and flatness problems. The slow-roll regime contributes most to the number of $e$-folds in the whole inflation. After the slow-roll regime, there will be a kinetic domination regime, and since now the potential decays and becomes gradually negligible. The evolution equation then approximately becomes:
\begin{equation}\label{EOM3}
  (F_X+2XF_{XX})\dot X+6HF_X(1+r)X=0.
\end{equation}

The theory of the warm little inflaton is the most efficient so far to realize warm inflation \cite{Berera2016}. As the theory of the warm little inflaton indicates, the dissipative strength increasingly weakens after the slow-roll regime, and the thermal dissipative effect finally becomes negligible after the end of inflation \cite{Berera2016,Rosa2019}. The radiation was produced constantly accompanied by inflation and its energy density can finally take over the energy density of the inflaton through a graceful transition without a separate reheating period. The evolution equation of the noncanonical field finally becomes
\begin{equation}\label{EOM4}
  (F_X+2XF_{XX})\dot X+6HF_XX=0.
\end{equation}
The above differential equation can be integrated exactly, for arbitrary $F$, to give the solution
\begin{equation}\label{solution}
  \sqrt{X}F_X=ka^{-3},
\end{equation}
where $k$ is a constant of integration.
It follows from the above equation that
\begin{equation}\label{solution1}
  X=\frac14\frac{L^2}{K^2}-\frac{L}{K}ca^{-3}+c^2a^{-6},
\end{equation}
where we write $c=k/K$, which is a constant, for convenience.
The energy density is then calculated as
\begin{equation}\label{rho2}
  \rho=\frac14\frac{L^2}K-cLa^{-3}+c^2Ka^{-6}.
\end{equation}
Using the result that we have obtained, we can describe the subsequent evolution of the Universe as follows. During the initial epoch of kinetic domination, the third term in Eq. (\ref{rho2}) is dominant. However, that term becomes small quickly relative to the radiation term $\sim a^{-4}$ that produced constantly during warm inflationary epoch (we have not expressed it explicitly here), and a radiation-dominated period of the Universe then ensues. The second term in Eq. (\ref{rho2}) gains prominence in the period of matter domination, which can be identified as dark matter. However, as the Universe evolves toward the present era, the first term begins to dominate and behaves like an observation-favoured cosmological constant, giving rise to the observed accelerated expansion of the Universe. The constant parameters $K$ and $L$ can be restricted by observations.

Combining the above equation with the expression for the pressure of the field, we get the state parameter as
\begin{eqnarray}\label{w}
  w&=&\frac{KX+LX^{\frac12}-\frac12m^2\phi^2}{KX+\frac12m^2\phi^2} \nonumber \\&=&\frac{-\frac14\frac{L^2}{K}+c^2Ka^{-6}-\frac12m^2\phi^2}{\frac14\frac{L^2}{K}-Lca^{-3}+c^2Ka^{-6}+\frac12m^2\phi^2}.
\end{eqnarray}
The state parameter takes different values for different epochs: $w\simeq-1$ during the slow-roll inflation epoch, $w\simeq1$ after the end of inflation and before radiation domination, $w\simeq0$ during matter domination, and $w\rightarrow-1$ as $a\rightarrow\infty$. The speed of sound is obtained as $c_s^2=1+\frac{1}{\frac{2cK}{L}a^{-3}-1}$. It is found that accompanying the expansion of the Universe, the sound speed decreases and approaches zero as $a\rightarrow\infty$.

This unified model successfully gives rise to the primary features of early inflation and reproduces a matter as well as a dark energy component in the later evolution of the universe. We now analyse the observational predictions of our model to see whether it is favored by observations. We first discuss the inflationary dynamics of the early universe. According to the calculations on the cosmological perturbations in general noncanonical warm inflation we have made in \cite{Zhang2018}, the scalar power spectrum in this concrete model is reduced to:
\begin{equation}\label{power}
  \mathcal{P}_R=\frac{9\sqrt{3}H^5TF_X(1+r)^{\frac52}}{2\pi^2m^4\phi^2}.
\end{equation}
Cosmological microwave background observations provide a good normalization of the scalar power spectrum $\mathcal{P}_R\approx10^{-9}$ on large scales. It can be founded from the above result Eq. (\ref{power}) that, compared to cold inflation or canonical warm inflation, the energy scale when horizon crossing is depressed by the noncanonical effect and thermal effect. This is good news to the assumption that the very early inflation can be described well by effective field theory.

The spectral index of scalar power spectrum is $n_s-1=\frac{d\ln\mathcal{P}_R}{d\ln k}\simeq\frac{\mathcal{\dot P}_R} {H\mathcal{P}_R}$,
which can be expressed as
\begin{eqnarray}\label{index1}
  n_s-1&=&\left[\frac{5c-16}{4-c}+\frac{6r}{(4-c)(1+r)}\right]\frac{\epsilon}{F_X(1+r)}+\frac{2\eta}
  {F_X(1+r)}\nonumber \\ &-&\frac{10r+4}{(4-c)(1+r)}\frac{\beta}{F_X(1+r)}-
  \frac{3r}{2(1+r)}\left(\frac{1}{c_s^2}-1\right)\delta \nonumber\\ &+&\frac{5cr+2r+2c+2}{2(4-c)r}b+
  \frac1{4-c}\frac{\epsilon\beta}{F_X(1+r)}
\end{eqnarray}
in the noncanonical warm inflation. Besides the slow-roll parameters $\epsilon$, $\eta$, $\beta$, $b$ and $c$ mentioned above, $\delta=\frac{\ddot\phi}{H\dot\phi}$ is also a slow-roll parameter which is much less than one. Guaranteed by the slow-roll conditions in noncanonical warm inflation \cite{Zhang2021,Zhang2014}, the spectral index, of order $\frac{\epsilon}{\mathcal{L}_X(1+r)}$, is found to be much less than one. So a nearly scale-invariant power spectrum is obtained, which is consistent with observations.

As is well known, the tensor perturbations do not couple to the thermal background, the gravitational waves are only generated by the quantum fluctuations as in standard inflation: $\mathcal{P}_T=\frac2{M_p^2}\left(\frac H{2\pi}\right)^2$.
The tensor-to-scalar ratio $R$ is then obtained:
\begin{equation}\label{ratio}
  R=\frac{\mathcal{P}_T}{\mathcal{P}_R}=\frac HT\frac{2\epsilon}{\sqrt{3}\mathcal{L}_X(1+r)^{\frac32}r}.
\end{equation}
As the scalar power spectrum is fixed by observations, the tensor perturbations generated in our noncanonical warm inflationary model is weaker than canonical warm inflation and standard inflation, which is safely allowed by observations. This characteristic is due to both the noncanonical effect and thermal effect in noncanonical warm inflation.

We now concentrate on the observational constraints of the model in the later evolutions. Using the current observed value of the cosmological constant, we have
\begin{equation}\label{lambda}
  \rho_{\Lambda}=\frac14\frac{L^2}{K}\simeq10^{-48}(GeV)^4.
\end{equation}
And as the observations suggests, the density parameter for dark energy at present epoch is $\Omega_{\Lambda}\simeq0.68$, while the matter density parameter at present epoch is $\Omega_{m}\simeq0.32$, so we can get
\begin{equation}\label{DEDM}
  \frac14\frac{L^2}{K}\approx-2cLa_0^{-3},
\end{equation}
where the subscript "0" refers to the present epoch. Since the third term in Eq. (\ref{rho2}) is absolutely negligible in later universe, the state parameter reduces to
\begin{equation}\label{w11}
  w=\frac{-\frac14\frac{L^2}{K}}{\frac14\frac{L^2}{K}-Lca_0^{-3}(1+z)^3},
\end{equation}
in post inflationary epoch, where $z$ is the redshift. From above equation, we have $w_0\simeq-0.68$ when $z=0$, which indicates the present universe undergoes accelerated expansion. Also, we can find out the value of redshift at which the universe started its transition from matter dominated decelerated epoch to its presently accelerated epoch. As is well known, cosmic accelerated expansion requires the state parameter $w\leq-\frac13$, so we obtain that the accelerated expansion begins at $z_{acc}\approx0.62$. The value we get is around a redshift of $z=\mathcal{O}(1)$, which is quite compatible with current observations \cite{PLANCK1}.

\section{\label{sec:level4}A $k$-essence unified model with a Lagrangian density having coupled kinetic and potential terms}

We now consider a special $k$-essence model having a Lagrangian density of the form $\mathcal{L}=F(X)V(\phi)$. The kinetic and potential terms in the expression of the Lagrangian density are
\begin{equation}\label{FX}
  F(X)=\frac{K}{M_p^4}X+\frac{L}{M_p^2}X^{\frac12},
\end{equation}
and
\begin{equation}\label{Vphi}
  V(\phi)=V_0\left[1+\left(\frac{\phi}{\mu}\right)^n\right],
\end{equation}
respectively, where $K>0$ and $L<0$, the parameters $K$ and $L$ are dimensionless, and the parameter $\mu$ has the dimension of mass. The potential in our model is a hybrid inflation potential \cite{Bassett2006}. As usual, the scalar field in this model has the dimension of mass; i.e., $[\phi]=[M]$. The kinetic term $F(X)$ is thus dimensionless, and the variable $X$ and the potential term $V(\phi)$ thus have the dimension of $[M]^4$.

The energy density and pressure of the inflaton field are obtained as
\begin{equation}\label{rho5}
  \rho_{\phi}=(2F_XX-F)V=V_0\left[1+\left(\frac{\phi}{\mu}\right)^n\right]\frac{K}{M_p^4}X,
\end{equation}
\begin{equation}\label{p5}
  p_{\phi}=\mathcal{L}=V_0\left[1+\left(\frac{\phi}{\mu}\right)^n\right]\left[\frac{K}{M_p^4}X+\frac{L}{M_p^2}X^{\frac12}\right].
\end{equation}
The state parameter is
\begin{equation}\label{w3}
  w=\frac{p}{\rho}=\frac{F}{2XF_X-F},
\end{equation}
which reduces to
\begin{equation}\label{w4}
  w=\frac{X+M_p^2LX^{\frac12}/K}{X}
\end{equation}
in our model.
The adiabatic speed of sound of the coupled noncanonical field describing perturbation travel is given by
\begin{eqnarray}\label{cs2}
  c_s^2&=&\frac{\partial p/\partial X}{\partial\rho/\partial X}=\frac{F_X}{2XF_{XX}+F_X} \nonumber\\
  &=&1+\frac12\frac{L}{K}M_p^2X^{-\frac12}<1.
\end{eqnarray}
The warm inflationary Universe is a multi-component system with nonzero temperature owing to the thermal dissipation of the inflaton field to the radiation.
The evolution equation for the $k$-essence scalar field in the warm inflationary scenario is now obtained as
\begin{equation}\label{EOM5}
  V(2XF_{XX}+F_X)\ddot\phi+3HF_XV\dot\phi+\Gamma\dot\phi+F_XV_{\phi}\dot\phi^2-FV_{\phi}=0.
\end{equation}
We then rewrite the motion equation above in terms of the variable $X$:
\begin{equation}\label{EOM6}
  (2XF_{XX}+F_X)\dot X+6HF_X X+2\frac{\Gamma}{V}X+\frac{\dot V}{V}(2XF_X-F)=0.
\end{equation}

The radiation production equation in the warm inflationary scenario is
\begin{equation}\label{rhor}
  \dot\rho_r+4H\rho_r=\Gamma\dot\phi^2=2\Gamma X.
\end{equation}
In the warm slow-roll regime, the production of radiation is quasi-static, and we thus have $4H\rho_r=\Gamma\dot\phi^2=2\Gamma X$. During the inflation, the dominant components of the Universe are the inflaton field and the radiation, and as guaranteed by the stability analysis \cite{Ian2008,Zhang2014,Zhang2021}, the total energy density $\rho$ and pressure $p$ can be written as
\begin{equation}\label{rhototal}
  \rho=\rho_{\phi}+\rho_r=V_0\left[1+\left(\frac{\phi}{\mu}\right)^n\right]\frac{K}{M_p^4}X+\frac{\Gamma X}{2H},
\end{equation}
\begin{equation}\label{ptotal}
  p=p_{\phi}+p_r=V_0\left[1+\left(\frac{\phi}{\mu}\right)^n\right]\left[\frac{K}{M_p^4}X+\frac{L}{M_p^2}X^{\frac12}\right]+
  \frac{\Gamma X}{6H}.
\end{equation}

The energy conservation equation states that $\dot\rho+3H(\rho+p)=0$, giving the fixed points $X_0=\frac14(\frac{K}{LM_p^2}+\frac{rM_p^2}{LV})^{-2}$. Evidently, $\rho$ decreases with time when $\rho>-p$ and increases when $\rho<-p$, indicating that any point corresponding to $\rho=-p$ is an attractor and, as is well known, will lead to exponential inflation.
In fact, $X_0$ corresponds to an instantaneous attractive fixed point and $X$ evolves slowly away from that point.
This can be seen as an analog of ``slow-roll'' potential-driven inflation in which the potential dominates the kinetic term and evolves slowly. Hence, in a similar way, we refer to the values of $\rho$, $p$ and $X_0$ calculated above as the slow-roll values. This scenario can simply be viewed as a type of warm $k$-inflation \cite{Peng2016}.
For slow-roll inflation, the Hubble parameter can be obtained: $H_0^2=-\frac{F(X_0)V}{3M_p^2}$. Under the slow-roll approximations of warm $k$-inflation \cite{Peng2016}, we can then attempt to get the number of $e$-folds:
\begin{eqnarray}\label{N}
  N&=&\int Hdt=-\int\frac{H_0}{\sqrt{2X_0}}d\phi \nonumber\\
  &=&-\frac{\sqrt{KV_0}}{\sqrt{6}M_P^3}\int_{\phi_i}^{\phi_e}\left[1+\left(\frac{\phi}{\mu}\right)^n\right]d\phi,
\end{eqnarray}
where $\phi_i$ and $\phi_e$ refer to the initial and final field values of the inflation, respectively.

As well as $\phi_i\gg\phi_e$, $N\simeq60$ can certainly hold, and our warm $k$-inflation is thus sustained long enough to solve the horizon and flatness problems. The slow-roll regime contributes most to the number of $e$-folds throughout the inflation.
In the post slow-roll inflationary epoch, $X$ moves away form the fixed point $X_0$, and the slow-roll approximation fails gradually. We write $X=X_0+\delta X$. Retaining the terms up to the first order in $\delta X$ from the evolution equation of $X$, we have
\begin{equation}\label{deltaX}
  \frac{\delta X}{X_0}=\frac{2\sqrt{2}M_p^2n(\frac{\phi}{\mu})^{n-1}F(X_0)}{3H_0L(1+r)+\sqrt{2X_0}Ln(\frac{\phi}{\mu})^{n-1}}.
\end{equation}
Inflation ends when $\frac{\delta X}{X_0}\sim1$. After inflation ends completely, the dissipation effect becomes extremely weak; i.e., $\Gamma\rightarrow0$. Additionally, with the progression of inflation, the inflaton flows down the potential, and we finally have $\frac{\dot V}{V}\rightarrow0$. The evolution equation can then be approximately written as
\begin{equation}\label{EOM7}
 (F_X+2XF_{XX})\dot X+6HF_X X=0.
\end{equation}
The above equation can be integrated into the analytic solution:
\begin{equation}\label{solution3}
  \sqrt{X}F_X=M_p^{-2}ka^{-3},
\end{equation}
where $k$ is a dimensionless integration constant. Using the above solution, we get the expression for the evolution of $X$:
\begin{equation}\label{X3}
  X=\frac{M_p^4}{K^2}(ka^{-3}-\frac12L)^2.
\end{equation}
Thus, the energy density in this model becomes
\begin{equation}\label{rho9}
  \rho=\frac{V_0L^2}{4K}-\frac{kV_0L}{K}a^{-3}+\frac{k^2V_0}{K}a^{-6}.
\end{equation}
Few remarks can be made for the above equation. During the fast-roll epoch after the slow-roll epoch, the third term is dominant, but it decreases quickly compared to the radiation term $\sim a^{-4}$ (which is produced constantly during the warm inflationary epoch and we have not expressed it explicitly here). Therefore, after a transient fast-roll epoch, a radiation-dominating period ensues. The second term in the above equation plays an important role in the period of matter domination, which can be identified by the presence of dark matter. As the Universe evolves toward the present era, the first term begins to dominate and behaves like a cosmological constant, driving the observed accelerated expansion of the Universe.
The expression for the evolution of the state parameter is now obtained as
\begin{equation}\label{w9}
  w=\frac{-\frac{L^2}{4}+k^2a^{-6}}{\frac{L^2}{4}-kLa^{-3}+k^2a^{-6}}.
\end{equation}
The state parameter $w\simeq-1$ during the slow-roll inflationary regime, and in the post slow-roll regime, there is a momentary fast-roll regime after inflation with $w\simeq1$. We have $w\simeq0$ when there is matter domination, and $w\rightarrow-1$ as $a\rightarrow\infty$.
The expression for the speed of sound is $c_s^2=1-\frac{1}{1-\frac{2K}{L}a^{-3}}$. We thus find that the speed of sound decreases towards zero as the Universe expands.

This model also successfully gives rise to the primary features of early inflation and reproduces a matter as well as a dark energy component in the later evolution of the universe. We then analyse the observational predictions of this model to see whether it is consistent with observations. We first discuss the inflationary dynamics of the early universe.
The early inflationary picture in this model can be seen as a type of warm $k$-inflation, and according to our previous work in \cite{Peng2018}, its power spectrum of scalar perturbations can be obtained as:
\begin{eqnarray}\label{5scalar}
  \mathcal{P}_{\mathcal{R}_{warm}}
  & \simeq & \frac{2\sqrt{3}}{c_s} \frac{T}{H} \Bigg(\frac{H}{\dot{\phi}}\Bigg)^2 \Bigg(\frac{H}{2\pi}\Bigg)^2 \nonumber \\
  & = & \frac{2\sqrt{3}}{c_s} \frac{T}{H} \mathcal{P}_{\mathcal{R}_{cold}},
\end{eqnarray}
where the result $\mathcal{P}_{\mathcal{R}_{cold}}=\Big(\frac{H}{\dot{\phi}}\Big)^2 \Big(\frac{H}{2\pi}\Big)^2$ is used in above equation. In warm $k$-inflationary regime, $T>H$ and $c_s<1$, so we have $\mathcal{P}_{\mathcal{R}_{warm}}>\mathcal{P}_{\mathcal{R}_{cold}}$, which can result in that the tensor-to-scalar ratio in this model becomes smaller than that in cold inflationary models.
According to the formula $n_s-1 = \frac{d \ln \mathcal{P}_{\mathcal{R}}}{d \ln k}$, the spectral index of scalar perturbations is obtained:
\begin{equation}\label{5index}
  n_s - 1 \simeq  \left( \frac{1}{4} b' - \frac{17}{4} \epsilon' - \frac{3}{2} \eta'- 3 \beta - \chi \right).
\end{equation}
In above equation, $\epsilon'$, $\eta'$ and $b'$ are slow-roll parameters in warm $k$-inflation \cite{Peng2016,Peng2018}. Through systemic stability analysis, it is found that they are all much less than unity within the inflationary period \cite{Peng2016}. And the parameters $\beta\equiv \frac{\ddot{\phi}}{H\dot{\phi}}, \chi\equiv \frac{\dot{c_s}}{c_s H}$ in Eq.(\ref{5index}) are also much less than unity during the inflationary period \cite{Mukhanov1999}. Therefore, the spectral index of scalar perturbations is nearly scale invariant, i.e. $n_s \simeq 1$, which is consistent with the observable result \cite{PLANCK2}.

Since tensor perturbations do not couple to the thermal background and primordial gravitational waves are only generated by the quantum fluctuations of inflaton field as in cold inflation, the spectrum of tensor perturbations are given by usual expression: $\mathcal{P}_{T_{warm}}=\frac{2}{M_p^2} \left( \frac{H}{2\pi} \right)^2$.
The tensor-to-scalar ratio $R$ in our model is thus expressed as:
\begin{equation}\label{5ratio}
  R=\frac{\mathcal{P}_{T_{warm}}}{\mathcal{P}_{\mathcal{R}_{warm}}}=\frac{8\sqrt{3}}{3}\frac{H}{T}c_s\epsilon.
\end{equation}
We can see that the tensor-to-scalar ratio $R$ in this model is smaller than that in cold inflation, which is more likely to be below the observational upper limit of $R$ ($R<0.06$) \cite{PLANCK2}. Therefore, warm $k$-inflation may be easier to consist with the observable results.

We then concentrate on the observational constraints of this model in the later evolutions. Using the current observed value of the cosmological constant, we have
\begin{equation}\label{lambda1}
  \rho_{\Lambda}=\frac{V_0L^2}{4K}\simeq10^{-48}(GeV)^4.
\end{equation}
As the observations suggests, the density parameter for dark energy at present epoch is $\Omega_{\Lambda}\simeq0.68$, while the matter density parameter at present epoch is $\Omega_{m}\simeq0.32$, so we can get the relation:
\begin{equation}\label{DEDM1}
  \frac{V_0L^2}{4K}\approx-2\frac{kV_0L}{K}a_0^{-3},
\end{equation}
where the subscript "0" also refers to the present epoch. Since the third term in Eq. (\ref{rho9}) is absolutely negligible in later universe, the state parameter can be reduced to
\begin{equation}\label{w22}
  w=\frac{-\frac{L^2}{4}}{\frac{L^2}{4}-kLa_0^{-3}(1+z)^3},
\end{equation}
in late time evolution of the universe. From above equation, we get $w_0\simeq-0.68<-\frac13$ when $z=0$, which indicates the present universe undergoes accelerated expansion. Also, we can find out the value of redshift at which the universe started its transition from matter dominated decelerated epoch to its presently accelerated epoch. As is well known, cosmic accelerated expansion requires the state parameter $w\leq-\frac13$. We then can obtain that the accelerated expansion begins nearly at $z_{acc}\approx0.62$, which is quite compatible with observations.

\section{\label{sec:level5}Conclusions}

We briefly introduced the warm inflationary scenario. Warm inflation generalizes the scope of inflation and does not require a separate reheating period. We analyzed two types of noncanonical field model to study the possibility of producing warm inflation in the early Universe and subsequently generating both dark matter and dark energy during later evolution in the appropriate order. We first presented a scalar field including a quadratic potential to achieve this unification. In the first model, the field is a noncanonical field with the Lagrangian density having separable kinetic and potential terms. The potential is dominant in the slow-roll inflationary regime, producing enough $e$-folds like other inflation models. The field first acts as a warm inflaton, leading to quasi-exponential expansion in the early universe and its dissipation to radiation constantly through the thermal damping term along with inflation. When the slow-roll regime ends, the potential energy is dominated by the kinetic energy, and the field energy density thus decreases sharply as $a^{-6}$ but does not disappear. From the exit of the slow-roll regime to the end of inflation, the dissipation effect continually weakens, and the dissipation strength $r$ thus decreases and finally approximates to zero. Radiation energy then dominates inflaton energy, and the inflation exits gracefully without a separate reheating period. The relic field behaves effectively as purely kinetic $k$-essence at late times. The analytical solution to the kinetic term $X$ can be obtained in the ad hoc model. Expressions for the energy density and state parameter in terms of the scale factor $a$ and for the adiabatic sound speed were obtained. The resultant energy density contained terms that unified dark matter and dark energy perfectly. A cosmological constant is quite favored by current observations as the source of dark energy \cite{PLANCK2}. Our model correctly reproduces the cosmological constant at late times. The adiabatic speed of sound approximates gradually to zero because the speed of sound decreases further as the scale factor increases, and there are thus no problems for structure formation. The cosmological perturbations of the early warm inflation and observational constraints of the model parameters in late evolution of the universe were calculated, which suggest the model is allowed by the observations. The unification of warm inflation, dark matter and dark energy is then realized in the first-type noncanonical field.

The second unified model is a noncanonical scalar field with a Lagrangian density having coupled kinetic and potential terms. The field has the dimension of mass in this type of model. In contrast with the first model, the slow-roll regime of inflation is driven by the kinetic term at the fixed point $X_0$. Meanwhile, in the first model, slow-roll inflation is mainly driven by the potential. In the slow-roll inflationary epoch, the field remain approximately at the attractor, and the radiation is produced constantly with the accelerated expansion, which can be viewed as a kind of warm $k$-inflation. In the post slow-roll regime, the field moves away from the inflationary attractor gradually. Once inflation ends completely, the thermal dissipation effect is near zero, and the variation in the potential is negligible. The equation for the evolution of the field can then be integrated analytically. Expressions for the variable $X$, universal energy density $\rho$, the state parameter $w$ and speed of sound $c_s$ at later times can thus be obtained analytically. The resultant energy density also contains terms that unify dark matter and dark energy perfectly. Additionally, a cosmological constant at late times, as favored by current observations \cite{PLANCK2}, is reproduced in the second model. The adiabatic speed of sound approximates gradually to zero as $a\rightarrow\infty$, because the speed of sound decreases further as the scale factor increases. We then analyse the cosmological perturbations of the early warm inflation and the observational constraints on the model parameters in late evolution of the universe. Relevant calculations indicate that the second model is also allowed by the observations. The unification of warm inflation, dark matter and dark energy is thus realized again in the second type of noncanonical field. We will consider other interesting models that can unify warm inflation, dark matter and dark energy in our future work.

%----------------------------------------------------------------------------------------------------
\acknowledgments This work was supported by the National Natural Science Foundation of China (Grant No. 11605100 and No. 11975132) and the Shandong Provincial Natural Science Foundation, China ZR2019YQ01.
%---------------------------------------------------------------------------------------------------------


\begin{thebibliography}{References}

\bibitem{Guth1981} A. H. Guth, Phys. Rev. D {\bf23}, 347 (1981).

\bibitem{Linde1982} A. D. Linde, Phys. Lett. B {\bf108}, 389 (1982).

\bibitem{Bassett2006} B. A. Bassett, S. Tsujikawa and D. Wands, Rev. Mod. Phys. {\bf78}, 537 (2006).

\bibitem{BereraFang} A. Berera, Phys. Rev. Lett. {\bf75}, 3218 (1995).

\bibitem{Berera1995} A. Berera and L. Z. Fang, Phys. Rev. Lett. {\bf74}, 1912 (1995).

\bibitem{Ian2008}I.G. Moss and C. Xiong, J. Cosmol. Astropart. Phys. 11 023 (2008).

\bibitem{Campo2010} S. del Campo R. Herrera D. Pav\'on and J.R. Villanueva, J. Cosmol. Astropart. Phys. 08 002 (2010).

\bibitem{ZhangZhu} X. M. Zhang and J. Y. Zhu, Phys. Rev. D {\bf87}, 043522 (2013).

\bibitem{Zhang2014} X. M. Zhang and J. Y. Zhu, Phys. Rev. D {\bf90}, 123519 (2014).

\bibitem{etaproblem}  M. Dine, L. Randall and S. Thomas, Phys. Rev. Lett. {\bf75}, 398 (1995).

\bibitem{etaproblem1} C. F. Kolda and J. March-Russell, Phys. Rev. D {\bf 60}, 023504 (1999).

\bibitem{Berera2006} A. Berera, Contemp. Phys. {\bf47}, 33 (2006).

\bibitem{BereraIanRamos} A. Berera, I. G. Moss and R. O. Ramos, Rep. Prog. Phys.{\bf72}, 026901 (2009).

\bibitem{Berera2000}A. Berera, Nucl. Phys. B {\bf585}, 666 (2000).

\bibitem{Lisa2004} L. M. H. Hall, I. G. Moss and A. Berera, Phys. Rev. D
{\bf69,} 083525 (2004).

\bibitem{Taylor2000} A. N. Taylor and A. Berera, Phys. Rev. D {\bf 62}, 083517 (2000).

\bibitem{Chris2009} C. Graham and I. G. Moss, J. Cosmol. Astropart. Phys. 07 013 (2009).

\bibitem{LiddleLyth}  A. R. Liddle and D. H. Lyth, {\it Cosmological Inflation and Large-Scale Structure} (Cambridge University Press, Cambridge, England, 2000).

\bibitem{PLANCK1} N. Aghanim \emph{et al}. (Planck Collaboration), Astron. Astrophys. {\bf641}, A1 (2020).

\bibitem{Scherrer2004} R. J. Scherrer, Phys. Rev. Lett. {\bf93}, 011301 (2004).

\bibitem{Bilic2009} N. Bilic, G. B. Tupper, and R. D. Viollier, Phys. Rev. D {\bf80}, 023515 (2009).

\bibitem{Brandenberger2021} R. Brandenberger, and J. Fr\"{o}hlich, J. Cosmol. Astropart. Phys. 04 (2021) 030.

\bibitem{quintessential}  P. J. E. Peebles and A. Vilenkin, Phys. Rev. D {\bf59}, 063505 (1999); K. Dimopoulos, Phys. Rev. D {\bf68}, 123506 (2003); M. Sami and V. Sahni, Phys. Rev. D {\bf70}, 083513 (2004); A. Gonzalez, T. Matos, and I. Quiros, Phys. Rev. D {\bf71}, 084029 (2005); I. P. Neupane and C. Scherer, J. Cosmol. Astropart. Phys. 05 (2008) 009.

\bibitem{Dimopoulos2019} K. Dimopoulosa and T. Markkanen, J. High Energy Phys. 01 (2019) 029.

\bibitem{Lidsey2002} J. E. Lidsey, T. Matos, and L. A. Ure\~{n}a-L\'{o}pez, Phys. Rev. D {\bf66}, 023514 (2002).

\bibitem{Zsembinszki2007} G. Zsembinszki, J. Phys. A: Math. Theor. 40 (2007) 5219.

\bibitem{Bose2009} N. Bose and A. S. Majumdar, Phys. Rev. D {\bf79}, 103517 (2009).

\bibitem{Nilok2009} N. Bose and A. S. Majumdar, Phys. Rev. D {\bf80}, 103508 (2009).

\bibitem{Liddle2008} A. R. Liddle, C. Pahud, and L. A. Ure\~{n}a-L\'{o}pez, Phys. Rev. D {\bf77}, 121301(R) (2008).

\bibitem{Sa2020} Paulo M. S\'{a}, Phys. Rev. D {\bf102}, 103519 (2020).

\bibitem{Arbey2021} A. Arbey and J.-F. Coupechoux, J. Cosmol. Astropart. Phys. 01 (2021) 033.

\bibitem{Berera1999} A. Berera, M. Gleiser and R. O. Ramos, Phys. Rev. Lett. {\bf83}, 264 (1999).

\bibitem{Herrera2006} R. Herrera, S. del Campo and C. Campuzano, J. Cosmol.
Astropart. Phys. 10 (2006) 009.

\bibitem{Xiao2011} K. Xiao and J. Y. Zhu, Phys. Lett. B {\bf 699}, 217 (2011).

\bibitem{MarGil2013} M. Bastero-Gil, A. Berera, R. O. Ramos, and J. G. Rosa, J. Cosmol. Astropart.
Phys. 01 (2013) 016.

\bibitem{MossXiong2006} I. G. Moss and C. Xiong, arXiv:hep-ph/0603266 (2006).

\bibitem{Franche2010} P. Franche, R. Gwyn, B. Underwood, and A. Wissanji, Phys. Rev. D {\bf81}, 123526 (2010).

\bibitem{Bean2008} R. Bean, D. J. H. Chung, and G. Geshnizjani, Phys. Rev. D {\bf78}, 023517 (2008).

\bibitem{Zhang2021} X. M. Zhang, A. Fu, K. Li, Q. Liu, P. C. Chu, H. Y. Ma, and J. Y. Zhu, Phys. Rev. D {\bf103}, 023511 (2021).

\bibitem{Berera2016} M. Bastero-Gil, A. Berera, R. O. Ramos, and J. G. Rosa, Phys. Rev. Lett. {\bf117}, 151301 (2016).

\bibitem{Rosa2019} J. G. Rosa and L. B. Ventura, Phys. Rev. Lett. {\bf122}, 161301 (2019).

\bibitem{Zhang2018} K. Li, X. M. Zhang, H. Y. Ma and J. Y. Zhu, Phys. Rev. D {\bf98}, 123528 (2018).

\bibitem{Peng2016} Z. P. Peng, J. N. Yu, J. Y. Zhu, and X. M. Zhang, Phys. Rev. D {\bf94}, 103531 (2016).

\bibitem{Peng2018} Z. P. Peng, J. N. Yu, X. M. Zhang, and J. Y. Zhu, Phys. Rev. D {\bf97}, 063523 (2018).

\bibitem{Mukhanov1999} C. Armend\texttt{$\acute{a}$}riz-Pic\texttt{$\acute{o}$}n, T.Damour, V. Mukhanov, Phys. Lett. B {\bf458} 209 (1999).

\bibitem{PLANCK2} N. Aghanim \emph{et al}. (Planck Collaboration), Astron. Astrophys. {\bf641}, A6 (2020).

\end{thebibliography}
\end{document}